\documentclass[letterpaper,twocolumn,10pt]{article}
\usepackage{usenix2025_SOUPS}
\usepackage{tikz}
\usepackage{amsmath}
\usepackage{filecontents}
\usepackage{url}
\usepackage{hyperref}
\usepackage{multirow}
\usepackage{pifont}
\usepackage{graphicx}
\usepackage{fmtcount}
\usepackage{array}
\usepackage{graphicx} 
\usepackage{booktabs}
\usepackage{enumitem}
\usepackage{makecell}
\usepackage{tikz}
\usepackage{colortbl}
\usepackage{xcolor}
\usepackage{titlesec}
\usepackage{mdframed}

\newcommand{\boldpartitle}[1]{\vspace{.03in}\noindent\textbf{#1.}}
\usepackage{mdframed}
\usepackage{tcolorbox}
\tcbset{
    summarybox/.style={
        colback=gray!5,        
        colframe=gray!40,    
        coltitle=black,     
        colbacktitle=gray!20,  
        fonttitle=\bfseries, 
        boxrule=0.6mm,        
        arc=2mm,               
        title={Research Question Overview},       
        width=\columnwidth,     
        top=0.5mm,               
        bottom=0.5mm,           
        left=1mm,            
        right=1mm
    }
}

\begin{document}

\date{}

\title{\Large \bf"Is it always watching? Is it always listening?" Exploring Contextual Privacy and Security Concerns Toward Domestic Social Robots}

\author{
{\rm Henry Bell}\\
Duke University
\and
{\rm Jabari Kwesi}\\
Duke University
\and
{\rm Hiba Laabadli}\\
Duke University
\and
{\rm Pardis Emami-Naeini}\\
Duke University
}
\def\plainauthor{Author name(s) for PDF metadata. Don't forget to anonymize for submission!}

\maketitle

\begin{abstract}
Equipped with artificial intelligence (AI) and advanced sensing capabilities, social robots are gaining interest among consumers in the United States. These robots seem like a natural evolution of traditional smart home devices. However, their extensive data collection capabilities, anthropomorphic features, and capacity to interact with their environment make social robots a more significant security and privacy threat. Increased risks include data linkage, unauthorized data sharing, and the physical safety of users and their homes. It is critical to investigate U.S. users’ security and privacy needs and concerns to guide the design of social robots while these devices are still in the early stages of commercialization in the U.S. market. Through 19 semi-structured interviews, we identified significant security and privacy concerns, highlighting the need for transparency, usability, and robust privacy controls to support adoption. For educational applications, participants worried most about misinformation, and in medical use cases, they worried about the reliability of these devices. Participants were also concerned with the data inference that social robots could enable. We found that participants expect tangible privacy controls, indicators of data collection, and context-appropriate functionality.
\end{abstract}
\section{Introduction} \label{sec:introduction}
Social robots are emerging as the next generation of Internet of Things (IoT) technology~\cite{nitto2017social, MISHRA2024123578}. These devices are characterized by their integration of advanced artificial intelligence (AI), physical embodiment, and ability to interact with users on a social level~\cite{naneva2020systematic, Henschel}. Social robots are gaining popularity in the United States for the utility they offer in various use cases~\cite{sobrepera2021perceived, de2019would, breazeal2011social}. They can support users' mental health~\cite{wada2007living}, offer companionship and assistance to patients in medical settings~\cite{jeong2015social, collins2024emotional,Carros2020}, and have been found to be beneficial for educational purposes~\cite{park2017growing, movellan2009sociable}. 

Although social robots possess many of the same capabilities as traditional IoT devices, they present unique privacy and security risks to users. To enable their functionality, social robots must collect a massive amount of sensitive and multi-modal data from users~\cite{tang2022confidant, tonkin2019privacy}. The AI-driven interactions of a social robot often prompt users to (over)share sensitive information~\cite{zhang2023s}, and if user data is used to train the AI models powering these robots, personal information could be leaked through memorization~\cite{bender2021dangers}. Since social robots are more mobile than traditional IoT devices~\cite{10.3389/frobt.2021.627958}, they can create additional threats to users' physical privacy~\cite{oruma2022systematic} by eavesdropping on conversations~\cite{denning2009spotlight}, or tampering with household objects~\cite{tadele2014safety, denning2009spotlight}. Despite these heightened risks, \textbf{limited research has examined U.S. users’ privacy and security awareness and contextual attitudes toward social robots, particularly within the home environment}.

Prior research, which has emphasized consumer interest in social robots despite their significant privacy and security concerns~\cite{Lutz}, primarily focuses on social robots within a single context, such as elderly adult~\cite{Carros2020,grabler2025privacy,shibata2011robot,wada2007living} or child users~\cite{tanaka2012children,kanda_robot_2004,serholt2016robots,dawe2019can,van2020teachers,han2005educational}. However, in a domestic setting, social robots are likely to engage with multiple users, and fulfill various different needs for each. Each interaction with a social robot presents a nuanced security and privacy landscape that warrants further exploration. This work expands on prior research by examining participants' attitudes, concerns, and expectations towards social robots in multiple scenario-based contexts. It is crucial to form a deep understanding of consumer risk awareness and expectations towards social robots now, while they are still in the early stages of commercialization. The present study addresses this need with three research questions:

\begin{itemize}[topsep=0pt,leftmargin=*, itemsep=-.5ex]
    \item \textbf{RQ1:} What level of knowledge and awareness do U.S.-based participants have about domestic social robots and their capabilities?
    \item \textbf{RQ2:} How are U.S.-based participants' security and privacy concerns or comfort levels with domestic social robots informed by different potential use cases?
    \item \textbf{RQ3:} What are U.S.-based participants' security and privacy expectations for domestic social robots in different potential use cases?
\end{itemize}

Through 19 semi-structured interviews with current users of smart home devices and AI chatbots, we took a critical step toward deepening the understanding of users' privacy needs in relation to domestic social robots. As users of similar technologies, our participants are well-positioned to recognize the dynamic and novel risks associated with social robots in domestic settings. To guide the future development of privacy-aware domestic social robots as they enter the U.S. market, we offer the following contributions:

\begin{enumerate}[topsep=0pt,leftmargin=*, itemsep=-.5ex]
    \item Through an in-depth qualitative analysis, we explore participants' understanding and awareness of risks associated with domestic social robots. Our findings reveal that participants conceptualize domestic social robots through their understanding of similar consumer technologies, such as smart home devices (e.g., smart speakers), and AI-enabled chatbots (e.g., ChatGPT). Additionally, the novelty of social robots makes identifying potential risks difficult for participants. Despite this, many expressed interest in adopting this technology.
    \item Our qualitative investigation found that participants' security and privacy concerns and expectations regarding domestic social robots varied widely based on the interaction context. While participants echo many concerns already surfaced in IoT literature, they expressed novel concerns about unique risks of social robots, such as threats to their physical, mental, and social privacy.
    \item We surface detailed and context-dependent privacy and security expectations. When used for a medical purpose, participants expected domestic social robots to be regulated under the Health Insurance Portability and Accountability Act (HIPAA); however, they currently are not. In scenarios where a social robot would be interacting with children, participants desired complete control over data collected, preferring the robot to request permission each time it would use data from a child.
\end{enumerate}
\section{Background and Related Work}\label{sec:relatedWork}
\boldpartitle{Despite having concerns, people are interested in IoT devices} Prior research has explored individual privacy concerns and mitigation strategies for IoT devices~\cite{Chow2015, Bhaskar2007}. Zeng et al.~\cite{Zeng2017} found smart home users had fewer security and privacy concerns when they trusted data-collecting entities (e.g., companies or governments) or did not feel personally targeted by advertisements. Emami-Naeini et al.~\cite{PEN2017} also showed that IoT privacy expectations are heavily influenced by the use context, with participants expressing more comfort when traditional privacy and data protection principles were applied. Despite privacy and security risks, users often prioritize convenience, cost-saving, and connectivity to adopt IoT devices for domestic use: Barbosa et al.~\cite{Barbosa2021} found that immediate benefits outweigh privacy concerns for many consumers. However, concerns grow when security-critical devices or ambiguous IoT applications are involved~\cite{Brush2011, Worthy2016, Choe2012}. Trust remains a key factor in user adoption~\cite{Brush2011, Worthy2016, Choe2012}, and Emami-Naeini et al.~\cite{emami2023consumers} found users are willing to pay premiums for enhanced security. 

\boldpartitle{Users have critical risk misconceptions toward AI chatbots} Chatbots powered by generative AI are designed to interact with users in a human-like way and are adaptable across various applications. Recent chatbots utilize deep learning architectures, including large language models (LLMs), to achieve high levels of performance~\cite{adamopoulou2020overview, zhan2023deceptive}. They can be used to generate persuasive messaging~\cite{karinshak2023working}, as an educational tool~\cite{iqbal2022exploring}, and to enhance text or produce code~\cite{skjuve2023user}.  

Prior work has established an extensive list of user privacy concerns related to conversational chatbots; these include concerns about data collection, storage, and usage~\cite{Dev2023, Agnihotri2023, Sannon2020, Griffin2021, Ai2020}, the disclosure of personal information~\cite{Fan2022, Gieselmann2022TheMC, Griffin2021, Saglam2021}, security~\cite{Kim2022, Marjerison22, ng2020simulating, gumusel2024user}, and transparency~\cite{Rodrigues20, Ai2020, cox_chatbotrefs}. Despite these concerns, users often disclose sensitive personal information to LLMs, even when warned against such disclosures~\cite{kshetri2023cybercrime}. It has also been well established that people anthropomorphize AI agents~\cite{mckee2023humans, sun_self-ai, Bi_Icreate}. There is evidence to suggest that users disclose more information and better adhere to product recommendations from chatbots perceived as more anthropomorphic~\cite{zhang2023s, ischen2020privacy}. Factors such as visual, conversational, or identity cues of humanness have all been shown to play a role in the anthropomorphism of chatbots~\cite{ischen2020privacy, go2019humanizing}. 

\boldpartitle{Social robots merge AI chatbots with sensor-heavy IoT devices, yet little research has explored users’ concerns and expectations toward these technologies} Recent research has examined the attitudes of different consumers toward social robots. Beer et al.~\cite{beer_older_2017} found that demonstrating robot capabilities improved older adults' acceptance of domestic assistive robots. Wada and Shibata~\cite{wada2007living} examined the long-term interaction effects of the therapeutic seal robot Paro on elderly individuals living in care facilities, reporting that regular social robot interaction could reduce feelings of loneliness and improve mood. Shibata and Wada~\cite{shibata2011robot} later demonstrated Paro's positive psychological effects on dementia patients, aligning with Moyle et al.~\cite{moyle2018care}, who found it encouraged social engagement.

In parallel with research on older adults, scholars have extensively studied how children interact with social robots in educational contexts. Pioneering work by Kanda et al.~\cite{kanda_robot_2004} introduced a humanoid robot into a Japanese elementary school setting. Over several weeks, children formed social bonds, treating the robot as a peer-like entity. Children remained positively engaged if the robot’s behaviors were socially contingent and responsive. Similarly, Tanaka and Matsuzoe~\cite{tanaka2012children} examined how a socially interactive robot influenced foreign language learning in children, observing that children displayed greater willingness to communicate with a robot that offered timely feedback and emotional support. These studies both align with the conclusions of Belpaeme et al.~\cite{belpaeme_social_2018}, whose review of child-robot interaction literature highlighted the value of personalization, adaptability, and social cues in sustaining children’s interest.

However, these technologies also raise privacy concerns. Gao et al.~\cite{gao2025consumer} analyzed YouTube and Bilibili reviews to study consumer acceptance of social home robots and found that privacy concerns were a factor influencing user intentions, accounting for 7\% of the comments. Another study surveyed U.S. adults (80\% of whom were parents) about their perceptions of privacy and attitudes toward social robots in the home~\cite{levinson2024surveying}. The results indicated that while participants recognized the utility of social robots in public areas such as living rooms, they were more concerned about the risks these devices pose to guests and children, who may be less aware of privacy implications.  In a co-learning workshop study by Levinson et al.~\cite{levinson2024our}, six families deliberated on allowing social robots to perform various privacy-sensitive tasks. Parents and children negotiated decisions based on specific use cases, highlighting the importance of context and use cases when designing social robots for multi-user home environments. 

Our study builds upon existing research on social robots by employing an interview-based approach to surface rich qualitative perspectives on the emerging privacy concerns participants have toward domestic social robots. While prior work has laregely focused on social robot acceptance and trust with specific populations, such as elderly users~\cite{Carros2020, Dereshev2019, wada2007living}, or parents and children~\cite{kanda_robot_2004, levinson2024our, levinson2024surveying}, there has been relatively little exploration of security and privacy perceptions toward social robots--particularly across a broader population. In addition, previous research has focused on social robots within specific domains, such as education or elderly assistive care, without systematically comparing concerns across these different applications. In contrast, our study investigates a wider range of use cases, considering contextual factors, user populations, and single vs. multi-use scenarios. By situating our findings within a security- and privacy-centric framework, we explore whether and how these privacy and security concerns align with or diverge from established privacy attitudes of other IoT technologies.
\section{Methodology} \label{methods}
To surface user privacy and security awareness, concerns, and expectations towards domestic social robots, we conducted 19 semi-structured interviews in March and April of 2024. We recruited U.S.-based participants from the Prolific crowdsourcing platform. No new insights emerged after interview 15, at which point we reached data saturation~\cite{saunders2018saturation}. Aligned with the guidelines to evaluate the reliability of data saturation, we interviewed 4 more participants~\cite{francis2010adequate}. We conducted the interviews online using the Zoom conferencing tool. We presented participants with an informed consent form at the beginning of each interview (see Appendix~\ref{interview_consent}). Our study protocol was approved by our institution's review board (IRB). We include the interview recruitment and procedure material in the Appendix~\ref{inteview_material}.

\boldpartitle{Participant recruitment}
We advertised our study as a research project to understand attitudes toward social robots. We explicitly did not mention privacy, security, or concern in the recruitment message to not prime participants about the goal of this work, mitigating demand characteristic bias~\cite{orne2009demand}. We recruited participants who were at least 18 years old and residing in the U.S.. To increase the quality of the responses, we only recruited Prolific participants with a task approval rating of at least 95\%. Prior to inviting the Prolific participants to join our interview study, we administered a screening survey to ask participants about their smart home device ownership, their experience using AI-enabled chatbots, and their demographic background. We then invited a sample of participants who had owned at least one smart home device and had used at least one AI-enabled chatbot. The interviews took, on average, 37 minutes (SD 9.5) to be completed, and participants were offered an incentive of \$20 for their participation. 

\boldpartitle{Social robot specification}
We designed a specification for a domestic social robot, which has similar capabilities to the social robots on the market. To increase the reliability of the study findings, we told participants that we were independent researchers who were assisting a social robot company in capturing consumers' feedback and opinions on the specifications. This was important to convince participants that the social robot we were discussing was a real device, rather than a hypothetical one. Our IRB office approved this use of deception. After the completion of all interviews, we sent participants a debriefing statement (see Appendix~\ref{interview_debrief}), where we explained the deceptive element and why we used the deception. We also gave participants the option to withdraw their data from being used in the study and raise any concerns with us and the IRB office. No participant expressed concerns or asked to withdraw from the study.

To build the social robot specification, we reviewed the domestic social robots available on Amazon under the searches ``Social Robot,'' ``Home Robot,'' and ``Domestic Robot'' as of February 2024. We chose these search terms as they are used interchangeably in the existing literature~\cite{de2015evaluation, rane2014study, han2005educational} and searched on Amazon since it is the largest e-retailer in the U.S.~\cite{altrad2021amazon}. For each selected term, we reviewed the search results until the presented products diverged from a domestic social robot. Since our goal was to evaluate participants' attitudes and concerns toward \emph{general-purpose domestic} social robots, we excluded any robot explicitly marketed for a single specific user group (e.g., children), with no reference of broader domestic use. Our final list consisted of five social robots: EBO X\footnote{\url{https://www.enabot.com/pages/ebo-x-family-robot-companion}}, Astro\footnote{\url{https://www.amazon.com/Introducing-Amazon-Astro/dp/B078NSDFSB}}, Eilik\footnote{\url{https://store.energizelab.com/products/eilik}}, Misa\footnote{\url{https://www.heymisa.com/}}, and Loona\footnote{\url{https://keyirobot.com/products/loona}}. Our goal was to design a \emph{desired} specification with the most comprehensive capabilities. To this end, we compiled a superset of features based on the specifications of the five final social robots. Our specification consisted of six features: 1) \texttt{visual recognition}, 2) \texttt{voice recognition}, 3) \texttt{expressive communication}, 4) \texttt{personalization}, 5) \texttt{navigation and mapping}, and 6) \texttt{internet connection}. The features and descriptions presented to participants are available in Table \ref{table:SRSpecification} in the Appendix. 

\boldpartitle{Interview scenario design}
\label{subsec: vignette des}
We constructed hypothetical scenarios to capture participants' concerns and opinions toward domestic social robots.  We walked participants through seven different scenarios, focusing on either the \textit{device recipient} or the \textit{purpose of use}. We considered four levels of \texttt{device recipient} which have been shown to impact security and privacy concerns: 1) \texttt{purchasing for self}~\cite{Lau2018}, 2) \texttt{purchasing for a child}~\cite{van2020teachers}, 3) \texttt{purchasing for a senior adult}~\cite{pal2018analyzing}, and 4) \texttt{purchasing for the household}~\cite{kramer2022empowering, stephenson2023s}. For the \texttt{purpose of use}, we included three levels: 1) \texttt{education}, 2) \texttt{medical}, and 3) \texttt{psychological therapy}. Prior research has explored the utility of social robots primarily for educational~\cite{van2020teachers}, medical~\cite{triantafyllidis2023social}, and therapy purposes~\cite{rakhymbayeva2021long, cruz2020social}. This is an example scenario that we presented to participants. This scenarios focuses on a \texttt{child} as the \texttt{device recipient}: 

\begin{quote}
Imagine that you are living in a family setting with a \texttt{child}. You are purchasing this specific social robot for the \texttt{child} to be the robot's primary user.
\end{quote}

\boldpartitle{Interview procedure}
Our interview consisted of three main sections (see Appendix~\ref{interview_script}): 1) knowledge and awareness toward social robots, 2) contextual privacy and security attitudes, and 3) privacy and security expectations toward social robots.

\emph{Section 1: Knowledge and Awareness Toward Social Robots.} In the second section, we asked questions to gauge participants' current preconceptions and awareness towards social robots. We then presented participants with a working definition for social robots, adapted primarily from the definition found on Wikipedia~\cite{enwiki:1219164302}:

\begin{center}
    \textit{``An artificial intelligence (AI) system that is designed to interact with humans and other robots by following social behaviors and rules attached to its role. Like other robots, a social robot is physically embodied.''}
\end{center}

We then shared our designed specification (outlined in Table~\ref{table:SRSpecification}) of the prototype social robot with participants, and asked them about their comfort and concern toward the device and its capabilities.

\emph{Section 2: Contextual Security and Privacy Attitudes.}
We then walked interviewees through the social robot scenarios and asked them to explain any specific comfort or concern they have towards each presented scenario.

\emph{Section 3: Security and Privacy Expectations Toward Social Robots.} Up until this point of the interview, we did not mention privacy or security so as not to bias participants. In this stage of the interview, for the first time, we mentioned security and privacy and asked participants to discuss what features and controls they expect social robots to have to address their concerns.

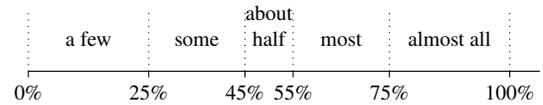
\begin{figure}[htbp]
    \centering
    \begin{tikzpicture}[x=1cm,y=1cm, scale=.8, transform shape, align=center]
        \draw[-] node[left]{}(0,0) -- (8.5,0) node[right] {};
        \foreach \x/\label in {0/0\%, 25/25\%,45/45\%,55/55\%,75/75\%, 100/100\%}
            \draw (\x/12.5,0) -- (\x/12.5,-0.1) node[below] {\label};

        \foreach \x/\label in {0/0\%, 25/25\%,45/45\%,55/55\%,75/75\%, 100/100\%}
            \draw[dotted] (\x/12.5,0) -- (\x/12.5,1);

        \foreach \y/\label in {12.5/a few,35/some,50/about\\half,65/most,87.5/almost all}
            \draw (\y/12.5,0.3) node[above] {\label};
    \end{tikzpicture}\vspace{-.15in}
    \caption{The terminology we use to report participants' percentage in \S\ref{sec:results}. Each term corresponds to a specific percentage range, enhancing the clarity and precision of reporting. For example, ``Most participants noted feeling comfortable with domestic social robots.'' means 55-75\% of interviewees reported as such.}\vspace{-.15in}
        \label{fig:terminology}
\end{figure}

\boldpartitle{Qualitative data analysis} 
We conducted a qualitative analysis of the collected data in two stages. The first stage included an iteration of structural coding where we categorized participants' responses using our three research questions. In the second-cycle coding, we applied thematic analysis~\cite{braun2006thematic} to surface the overarching themes and create main codes (e.g., purchase decision factors, expected features of a social robot). All the interviews were coded independently by two researchers on the team. The coders met periodically to jointly create the codebook and resolve any disagreements in the coding. Since the two researchers resolved all disagreements in the codes, inter-rater reliability was not calculated~\cite{10.1145/3359174}. Due to the qualitative nature and small sample size of our study, we adopt a terminology used in prior work~\cite{apthorpe2022you, PEN2019, habib2022identifying, habib2020s} and described in Figure~\ref{fig:terminology} to provide a quantitative representation of the frequency of participant responses.

\subsection{Limitations} \label{methods:limitations}
As social robots in the U.S. are still in the early stages of commercialization, most participants were unfamiliar with them prior to our interview. Future studies should investigate the concerns of long-term social robot users. Since we used Prolific's pre-screeners to recruit participants, we did not include people who may have used, but not owned, smart home devices. This could have excluded participants who could not afford smart home technology.  Additionally, our participant sample consisted mostly of highly educated people, all from the U.S., between the ages of 35 and 54 - there is value in exploring the perspectives of individuals who do not belong to this demographic group. Our discussion of both child and elderly users is only from the perspective of adults, not in either of those groups. The perceived benefits and concerns, specifically for elderly users, could be significantly different from those of adults imagining how an elderly user would use a social robot. To maintain the flow of the interview, we presented the device recipient and purpose of use scenarios sequentially; however, the study could be improved by randomizing the order in which each level and grouping of scenarios was shown to account for order effects~\cite{10.1145/2207676.2208589}. While the factors and levels we investigated in our scenarios are supported by previous research, they are not comprehensive and participants may have had unique concerns in other contexts. A future quantitative exploration could explore a wider range of scenarios and mitigate bias by controlling for the order in which they are presented. Lastly, participant responses could have been biased due to biases associated with self-assessment~\cite{karpen2018social}, social desirability~\cite{grimm2010social}, and the privacy paradox~\cite{zhang2023aiot}.
\section{Results} \label{sec:results}
Our interview sample consisted of 12 female and 7 male participants with an average age of 46. All participants had prior experience with both AI-enabled chatbots and smart home devices. We provide the demographic information of our participants as well as the technologies they have used in Table~\ref{tab:demographic} and Table~\ref{tab:technologies}, respectively.

\begin{table*}[t]
\centering
\resizebox{\textwidth}{!}{%
\footnotesize{
\begin{tabular}{lllll}
\toprule[1.1pt]
\textbf{Participant ID} & \textbf{Gender} & \textbf{Age} & \textbf{Ethnicity} & \textbf{Education} \\ \midrule
P1 & Female & 45-54 & Hispanic or Latino, or Spanish Origin of any race & Regular high school diploma \\
P2 & Male & 25-34 & White & Bachelor’s degree (e.g., BA, BS) \\
P3 & Female & 25-34 & White & 1 or more years of college credit, no degree \\
P4 & Female & 35-44 & White & Regular high school diploma \\
P5 & Female & 35-44 & White & Bachelor’s degree (e.g., BA, BS) \\
P6 & Male & 55-64 & White & Master’s degree (e.g., MA, MS, MEng, MEd, MSW, MBA) \\
P7 & Female & 45-54 & Hispanic or Latino, or Spanish Origin of any race + Brazilian & Bachelor’s degree (e.g., BA, BS) \\
P8 & Male & 55-64 & White & Associate's degree (e.g., AA, AS) \\
P9 & Female & 45-54 & White & Some college credit, but less than 1 year of college \\
P10 & Male & 45-54 & Asian + White & Bachelor’s degree (e.g., BA, BS) \\
P11 & Male & 35-44 & Black or African American & Associate's degree (e.g., AA, AS) \\
P12 & Female & 55-64 & Hispanic or Latino, or Spanish Origin of any race + White & Doctorate degree (e.g., PhD, EdD) \\
P13 & Female & 45-54 & Hispanic or Latino, or Spanish Origin of any race + White & Doctorate degree (e.g., PhD, EdD) \\
P14 & Female & 45-54 & White & Doctorate degree (e.g., PhD, EdD) \\
P15 & Male & 45-54 & White & Bachelor’s degree (e.g., BA, BS) \\
P16 & Male & 45-54 & Hispanic or Latino, or Spanish Origin of any race + White & Associate's degree (e.g., AA, AS) \\
P17 & Female & 35-44 & White & 1 or more years of college credit, no degree \\
P18 & Female & 45-54 & White & Regular high school diploma \\
P19 & Female & 35-44 & White & Professional degree beyond bachelor’s degree (e.g., MD, DDS, DVM, LLB, JD) \\
\bottomrule[1.1pt]
\end{tabular}
}
}\vspace{-.15in}
\caption{Participants' demographic information.}
\label{tab:demographic}\vspace{-.15in}
\end{table*}

\subsection{Knowledge Toward Social Robots}
\label{section:GeneralAwareness}
We began our interview by asking participants a few questions to capture their understanding and awareness of the term ``social robot.'' About half of the participants reported no familiarity with the term ``social robot''. We asked all participants to provide their best interpretation of a social robot. Most participants anchored their definition to an existing digital technology, commonly to available consumer technologies, such as Amazon's Alexa or OpenAI's ChatGPT. This suggests that for individuals for whom social robots are still novel, perceptions of this technology are closely shaped by knowledge of other technologies that share similar capabilities. Most participants defined a social robot as a digital system with capabilities for self-learning, internet connectivity, and social engagement.

The novelty of social robots sometimes made it hard for participants to anticipate the risks associated with them. P17 explained that she wouldn't be able to conceptualize concerns due to her lack of experience:
\begin{quote}I don't know. I wouldn't have any concerns… It's such a new concept. I wouldn't even know where to begin until [I experienced it].
\end{quote}
\begin{mdframed}[linewidth=2,linecolor=gray, topline=false,rightline=false,bottomline=false,innertopmargin=-.5pt,innerleftmargin=5pt,innerrightmargin=0pt,innerbottommargin=-.5pt]
    {\color{gray}\textbf{Summary of 4.1:}}
About half of the participants were not familiar with social robots before the interview. These participants tended to conceptualize social robots through their existing understanding of smart home devices and AI-enabled chatbots. Additionally, the novelty of social robots sometimes made it hard for participants to identify potential concerns with the technology.
\end{mdframed}

\subsection{Concerns Toward Social Robots}
\label{subsec:concerns}
We included six critical and potentially privacy-invasive features of a social robot in our designed specification (see \S\ref{methods}). We presented participants with the designed specification and asked them questions to capture their perceptions of the presented capabilities and their concerns toward the described social robot.

\boldpartitle{Participants were resigned to social robot privacy violations}
Some participants explained that while they were not necessarily comfortable with a social robot collecting their data, they viewed it as inevitable, a feeling known as privacy resignation~\cite{hoffmann2016privacy}. While P5 expressed discomfort with data collection, they were ultimately resigned to allow it:
\begin{quote}[it’s] so terrifying. If somebody laid it all out for me, like, here's everything that Meta knows. Here's everything that Apple and Google know, and how easily they can get that, I'd probably feel uncomfortable. But I think that is part of the social contract that we have with all these companies.
\end{quote}
\boldpartitle{Participants worried about their personal safety}
A few participants had concerns surrounding the personal safety threats of a social robot. P14 could imagine how a social robot could be used by others to threaten her safety:
\begin{quote}
It shouldn't be abusing and harassing me if I bought it, or serving as a vehicle through which I can be abused and harassed by others
\end{quote}
\boldpartitle{Mobility of the social robot was not a primary concern among participants}
Despite these concerns about personal safety, only a few participants expressed concern about the navigation and mapping capabilities of the social robot. While many participants did not focus on mobility as a concern, a few described scenarios where the social robot's mobility would increase the perceived threat of other privacy risks. 

P13 explained that her concern that the social robot could collect sensitive information was increased by its navigational capabilities: 
\begin{quote}The robot could go around and dig through my purse and look at my credit cards... it could be used for all sorts of things if it's in your home, and it can move around and map.
\end{quote}
\boldpartitle{Participants expressed significant concerns toward the passive and active collection of visual and voice data}
About half of the participants reported having privacy concerns regarding the audio and video collection of social robots, either being collected as a result of intentional interaction with the device or being passively collected without users' interaction. P5 was concerned about where identifiable information would be stored: 
\begin{quote}So as it's doing all this learning… like learning my face, learning my voice. Where is all [the data] being stored?
\end{quote}
Similarly, P6 discussed their privacy concerns with being in the presence of a social robot and expressed interest in having information about devices' privacy and security practices:
\begin{quote}What's going on with the data on the back-end? How much of my privacy am I giving up by just being in the environment with that robot?
\end{quote}
 No participants talked about unique risks introduced by the AI components of social robots. A key example is memorization, where the AI models powering these robots can unintentionally memorize sensitive information during training and potentially expose this data to other users~\cite{bender2021dangers}. 

 \boldpartitle{Participants were not comfortable with data inference}
 A few participants worried about the information that could be inferred from their interactions with a social robot. P4 described how she was uncomfortable with what Google inferred about her:
 \begin{quote}I have actually looked through the data that Google knows about me, and it's a little bit creepy because they know things that I have not directly told them... How did they get that information?
 \end{quote}
 When comparing a social robot to her current smart speaker, P4 explained that data inference from a social robot would be more concerning since it would collect more data:
 \begin{quote}[Social robots] could be a little smarter, and that could be a little more concerning, just because, you know, maybe they're inferring even more things... maybe because there would be more information that they would have.
 \end{quote}


Prior research has shown that the context in which the technology is being used can significantly impact users' security and privacy concerns and perceptions~\cite{sun2021child, pal2018analyzing}. We presented hypothetical data collection and use scenarios in which a household social robot is being primarily purchased for a \texttt{device recipient} (four levels: yourself, children, elderly, household) or to satisfy a \texttt{purpose} (three levels: educational, medical, therapy). 

\boldpartitle{Most participants did not worry about scenarios in which they were the primary user of a household social robot} 
Most respondents expressed little or no privacy or security concerns toward purchasing a social robot for themselves. The most commonly mentioned benefit of having a social robot was companionship. Other frequently mentioned benefits of social robots include safety and productivity. P1 discussed why they were comfortable with purchasing such a robot for themselves:
\begin{quote}It would make me feel safer, if anything were to happen, I’m sure I could ask it to dial 911.
\end{quote}
When imagining purchasing a social robot for themselves, a few participants reported having concerns about the data practices of the device. One participant in particular worried that their sensitive information could be used against them:
\begin{quote}I think the most uncomfortable aspect of it is like, how much does it know about me? Can that information be used by somebody else, and can it be used against me in any way?
\end{quote}
One expected use of a social robot in this scenario was to monitor relevant household items. P6 explained that a social robot would be beneficial for reminding him about the needs of his dog:
\begin{quote}I'm not necessarily gonna ask it to fill the water bowl, but at least inform me. `Hey no, he needs this, or it looks like he's low on treats, or it looks like he wants to go out', so I don't constantly have to be like getting up and doing those kinds of things.\end{quote}

\boldpartitle{Most participants were concerned about their children interacting with household social robots}
Most participants mentioned at least one privacy or security concern with a child being the primary user of a household social robot. There were significant concerns surrounding children's data security and privacy, and social development. P9 discussed their
significant concerns about their kids' natural behaviors being recorded by a social robot:
\begin{quote}The visual thing would bother me with [my daughter] being a child. I've had to be careful with her…that she's not trying to get changed or anything.\end{quote}
A few participants imagined a social robot being used to help their child complete chores. P14 expected the robot to remind their child to stay on task, and to help with some chores:
\begin{quote}If it could assist the child with their homework, or their tap their list of chores, or remind them to be doing things they need to be doing instead of Mom and Dad having to be that person [that would be] helpful... could it clean their room, or, you know, fold their laundry?
\end{quote}
While some participants mentioned companionship as a potential benefit of allowing a child to interact with a social robot, a few participants worried that this ``artificial'' socialization would have negative impacts on their children's development. P5 explained that she would prefer her children to interact with other children, rather than a social robot:
\begin{quote}I wouldn't buy my kid a robot friend. I would encourage my child to have social relationships with actual human children. I just don't think that is something I would layer in as an experience for my kid.
\end{quote}
A few participants worried that children might misuse the device to access age-inappropriate content online, or make purchases without informing their parents. P4 was particularly concerned about this:
\begin{quote}[I] also have concerns with the child accidentally buying things, or signing up for [subscriptions] that would cost money.
\end{quote}

\boldpartitle{Despite concerns toward being deceived, the perceived benefits of domestic social robots for senior adults outweighed the potential privacy harms} 
Most participants were comfortable with purchasing a social robot for an elderly family member. It is important to note, however, that most of our participants are under the age of 65 and would not be considered elderly. These participants often discussed the purchase of a social robot for their own elderly parents. This should be kept in mind when interpreting the results, as elderly users may feel differently if asked to purchase this robot for themselves~\cite{lee2018reframing, london2023ethical}. About half of the participants reported that a social robot would be beneficial in providing companionship for elderly family members. Assisting with medical care needs for senior adults was the second most frequently perceived benefit of social robots. Some participants felt that a social robot could bring them peace of mind. P6 described specifically how this could be a benefit:
\begin{quote}I could definitely see [the robot] as being a really good watchdog, so that if anything happened with my mother, we would receive some sort of a notification.
\end{quote}
Lack of usability was the most frequent concern when discussing elderly users of social robots. About half of the participants worried that the device would not be usable for an elderly user, and that it would be hard to convince an elderly family member to adopt the technology. P13 worried that an elderly user would struggle to use a social robot. A few participants mentioned having privacy and security concerns about senior adults' use of social robots. Our participants were primarily concerned that the robot would influence senior adults to share sensitive information about themselves. P7 said:
\begin{quote}I would want to make sure that their information is safe. Because, especially with the elderly, they don't always know when they're being swindled.\end{quote}

\boldpartitle{Maintaining confidentiality was the main privacy concern when sharing a social robot with household members} In a communal setting where multiple users would share a social robot, about half of the participants worried about their data being leaked to other users. P14 worried about potential malicious behavior in a shared living setting:
\begin{quote}Is there sensitive information that the other roommate could, if they had ill intention, access?
\end{quote}
A few participants mentioned bystander privacy as their main concern with social robots being shared among household members. P2 discussed the importance of ensuring that everyone in the communal setting is comfortable with a social robot: 
\begin{quote}I would be comfortable, but it might make [my roommate] a little uncomfortable.
\end{quote}

\boldpartitle{Misinformation was the primary concern in using social robots for education}
Participants saw potential educational benefits of using social robots, especially for children. These participants reported that social robots could be useful in helping children with their homework and assisting educators outside of a classroom setting. P9 described how her child attends school virtually, and so having a social robot to help teach content would be helpful:
\begin{quote}My daughter does cyber school… this is an everyday battle, and I'm like I didn't learn this stuff 30 years ago. So yeah, I would be absolutely comfortable.
\end{quote}
Some participants still worried about the reliability of social robots as an educational tool. Misinformation and potential hallucination effects were frequently mentioned. Prior research supports this concern, documenting hallucinated responses to questions across many topics~\cite{zhan2023deceptive}. P19 worried about information accuracy, comparing the social robot to ChatGPT and Alexa:
\begin{quote}I don't know that I would trust it cause I know it could be just like how ChatGPT works or how [Alexa] works. I know sometimes they come up with nonsense.
\end{quote}

\boldpartitle{Lack of reliability was the main concern for using social robots in a medical context} 
The most common concern participants had about using a social robot in a medical setting was the lack of reliability of the decisions made by the device. Despite this, about half of the participants reported being comfortable using a social robot as part of a larger care plan. P3 saw medical social robots as a starting point for treatment, but not a replacement:
\begin{quote}I think anything like that is open to misinformation, so I would use it as a starting point, but not to replace a doctor.
\end{quote}
Some participants worried about the medical data a social robot would collect. P19 explained she would be comfortable with a social robot doing some tasks, but not collecting data: 
\begin{quote}[Medical use] is a very wide ranging topic. If it was like reminding me to take pills, or to check my [pulse oxygen] or blood pressure, sure, great! If it was recording any of that info or giving any type of advice, absolutely not.
\end{quote}

\boldpartitle{The use of the social robots for therapy did not introduce new privacy or security concerns} 
Despite seeing convenience as a benefit, about half of the participants saw social robots as less effective than human therapists. These participants tended to believe that human-to-human interaction was necessary for effective therapy, and expected interacting with a social robot to be shallow in comparison. This belief is similar to the ``Perceived Loss of Human Touch'' often described with generative AI in medical contexts~\cite{singh2022quantifying}. P7 mentioned:
\begin{quote}I think it could be very helpful in certain scenarios, but again... is it helping the evaluation process? Or is it helping like a mental health patient who can use support in some way? I feel like that could be a supporting technology, assistive in some ways
\end{quote}

A few participants had concerns about how the unreliability of generative AI could harm people seeking psychological help. P19 worried about the impact that inaccurate information could have:
\begin{quote}I don't know the advice [social robots] are giving, and if someone is in a vulnerable state, that could be extremely dangerous and harmful.
\end{quote}
\begin{mdframed}[linewidth=2,linecolor=gray, topline=false,rightline=false,bottomline=false,innertopmargin=-.5pt,innerleftmargin=5pt,innerrightmargin=0pt,innerbottommargin=-.5pt]
    {\color{gray}\textbf{Summary of 4.2:}}
Participants’ concerns about social robots were shaped by their experiences with smart technologies, focusing on audio and video data collection while overlooking navigational risks. They were least concerned about owning a social robot, but were hesitant about purchasing one for a child. While recognizing benefits for elderly users, they noted usability challenges. Multi-user robots raised concerns about data leakage and bystander privacy. Participants favored low-stakes tasks but opposed replacing human practitioners in critical roles.
\end{mdframed}
\subsection{Transparency For Social Robots} \label{subsec:TransparencyExpectations}
Almost all participants expected social robots to provide trustworthy information about their security and data practices through multiple modalities. Aligned with prior work on IoT devices~\cite{norval2023room, chalhoub2021did, kulyk2020does}, our participants wanted to know what data is being collected and used for, who it is shared with, and how it is protected.

\boldpartitle{Information about what data is being collected and inferred} About half of our participants reported being interested in knowing what type of information the social robot can collect and learn about them. In addition, participants wanted to be informed about how this information is collected. 

\boldpartitle{Information about the purpose of data collection} About half of the participants expressed interest in knowing what the data collected on them would be used for. Consistent with prior research, participants were more accepting of data collection that personalized the devices or contributed to their functionality~\cite{Zheng2018}. P15 was not comfortable with sensitive data being collected, but was okay with personalization:
\begin{quote}I think this goes without saying [that I don't want the robot] giving personal health data or a bank account data or anything... But there's lots of stuff I don't mind. I mean, if it's personalizing the experience, I don't mind giving up a little bit.
\end{quote}
\boldpartitle{Information about how the data is being protected} Most participants felt that both the social robot manufacturer and the user of a social robot were responsible for maintaining security and privacy. About half of the participants wanted information on the data protection practices in place. Some wanted to know about the security measures of the device to ensure their robot could not be hacked and used by adversaries to collect data. P18 explained that she would want to know about the safeguards in place, as well as the manufacturer's reputation in securing user data:
\begin{quote}I would want to know what is going on with [my] information. How secure is the company that produces this robot?... How well does it do with security? Does it have a number of breaches? What protocols are put in place to protect sensitive information?... What's the company's track record? 
\end{quote}
\boldpartitle{Information about user privacy controls} Some participants wanted control over who could access the device and its collected information. P13 described a scenario in which access control would be valuable for keeping house guests from accessing the social robot:
\begin{quote}Say, for instance, I invite someone over to stay at my house that I may not know completely well, or have a relative come and stay at my house. I don't want them to be able to easily access [the robot]
\end{quote}

\boldpartitle{Information about the models that power the AI functionality of the social robot} When prompted about AI transparency specifically, about half of the participants wanted to have more information about the model powering the conversational AI of a social robot. This included the model that was being used, the data it was trained on, and its ability to adapt and learn from the user. A few participants also mentioned the trustworthiness of the AI. 

\boldpartitle{Providing transparency through multiple channels, including on package and through the device}
When asked about the preferred modality of transparency in social robots, most participants reported that they would like the security and privacy information to be presented to them through multiple channels as opposed to one single method. When explaining his preferred method of communication, P10 said:
\begin{quote}
In every way. Writing, video, disclaimers... I think clarity is kindness... it's the responsibility of the manufacturer to be upfront with all the possibilities.
\end{quote}
Most participants expressed a preference for security and privacy information to be on a device's packaging or to be available online. Participants noted that, while convenient, it might be infeasible to fit all of a device's information onto its physical packaging. P14 explained how she would expect the information to be present on the device's packaging, but would also seek out more details online.
\begin{quote}I think it should be on the box, and I definitely think there should be a website that details the specifics. I [would] probably look on the website honestly.
\end{quote}
A few participants reported that they prefer the social robot to communicate its own privacy and security information, expecting the robot to leverage its voice and visual interfaces to convey its security and data practices to the users. 
\begin{mdframed}[linewidth=2,linecolor=gray, topline=false,rightline=false,bottomline=false,innertopmargin=-.5pt,innerleftmargin=5pt,innerrightmargin=0pt,innerbottommargin=-.5pt]
    {\color{gray}\textbf{Summary of 4.3:}}
    Participant expectations of security and privacy transparency in social robots were similar to expectations towards traditional IoT and smart home technologies. Participants primarily wanted to know what data was being collected, what it was being used for, and who it was being shared with. They also wanted to know about the privacy controls available for social robots and the AI models that power them. Participants expected this information to be available in various modalities such as online, on the device packaging, or through the social robot's communication interface.
\end{mdframed}

\subsection{Expectations Toward Social Robots} We asked participants about the features they expect social robots to have to protect their security and privacy. 
\label{subsec: DesiredSNP}

\boldpartitle{On-device visual and audio cues to signal data collection and processing}
Some participants mentioned that a visual or audio cue that the device was currently collecting data would mitigate some of their privacy and security concerns. P4 worried that they would not be able to tell when a social robot was collecting information:
\begin{quote}Is it always watching? Is it always listening? Is there some sort of cue I would give it that would cause it to listen/see?
\end{quote}
\boldpartitle{Ability to review and delete the collected data}
A few participants expected social robots to behave similarly to some smart speakers~\cite{Lau2018}, letting users review and delete the data a social robot collects. When discussing their comfort with the speech recognition capabilities of social robots, P4 said:
\begin{quote}I do make a habit of going in and deleting all of the voice recordings, double checking that it's not hearing things.
\end{quote}
\boldpartitle{Being able to cover device sensors to limit data collection}
A few participants expected to be able to cover and/or unplug the sensors on social robots. P14 mentioned that she would want the ability to disable data collection with a physical button on the device:
\begin{quote}[I would want] a kill switch to not have anything be recorded... I'd want to have full control over what information is going out
\end{quote}
\boldpartitle{Parental control to manage data practices of social robot}
Parental controls were the most frequently requested feature among participants who reported being concerned about children's use of domestic social robots. Participants wanted to have control over what data a social robot could collect from their child. P11, for example, preferred the robot to ask for permission before using any visual data of their child:
\begin{quote}
Give me an authorization request... like `are you okay with using your kid's image'?... An authorization to consent to the use of my child's images.
\end{quote}
\boldpartitle{Policies and regulations to control the data practices of social robots} 
Some of our participants mentioned the need for having strong security and privacy regulations and standards for social robots, especially when being used in sensitive use cases, for example, assisting an elderly household member with their health needs. P7 mentioned:
\begin{quote}We have very strict rules about privacy and medicine. As long as it stays within [HIPAA] I think it could be very helpful.
\end{quote}
Unless provided by a covered entity, data collected by household social robots is not protected under health privacy laws, such as the Health Insurance Portability and Accountability Act (HIPAA). This response from participant 7 indicates a lack of awareness of the available regulations to protect users' data when interacting with domestic social robots.
\begin{mdframed}[linewidth=2,linecolor=gray, topline=false,rightline=false,bottomline=false,innertopmargin=-.5pt,innerleftmargin=5pt,innerrightmargin=0pt,innerbottommargin=-.5pt]
    {\color{gray}\textbf{Summary of 4.4:}} Participant expectations toward security and privacy controls in social robots closely mirrored preferences identified in prior research on smart home devices broadly. Participants wanted clear signals when data collection was happening and tangible controls to disable specific sensors based on the current context of use. They also wanted strict and granular parental controls and the ability to review, collect, and delete all data a social robot collects. Lastly, participants emphasized the importance of strong and clear regulations and standards for social robots.
\end{mdframed}
\section{Discussion}\label{sec:disc}
Our study contributes to the ongoing discourse on privacy and security concerns surrounding domestic social robots. While prior research has primarily focused on user acceptance and trust~\cite{gao2025consumer, Berzuk_SRExpectation, Kwon_SRExpectations, pena2020human, Zhu_TrustedListener}, our work investigates security and privacy attitudes toward these robots. Additionally, existing studies on social robots, including those addressing security and privacy, often examine predefined use cases such as education or assistive care~\cite{Carros2020, Dereshev2019, beer_older_2017, kanda_robot_2004, tanaka2012children}, or focus on specific user groups like children~\cite{levinson2024our} or parents~\cite{levinson2024surveying}. We explore concerns across multiple use cases, allowing for a comparative analysis of privacy and security risks. 

In brief, our findings reveal that most participants were unfamiliar with the term ``social robot,'' yet they consistently associated these devices with self-learning, social engagement, and internet connectivity (See \S \ref{section:GeneralAwareness})--aligning with prior literature~\cite{horstmann2019great}.  Nearly all participants voiced privacy concerns regarding domestic social robots even before being explicitly prompted (See \S \ref{subsec:concerns}), with primary concerns centered around the audio and visual data collection capabilities of these devices (See \S \ref{subsec:concerns}). While similar concerns exist for smart home devices~\cite{Zheng2018, Lee&Kobsa2017,chalhoub2021did}, our findings suggest that domestic social robots introduce an additional layer of apprehension due to their autonomous interactions; many worried about data collection occurring without their awareness or explicit consent (See \S \ref{subsec:concerns}). These concerns were highly context-dependent. Participants generally expressed fewer concerns about personal ownership of domestic social robots but were significantly more cautious about their use by children and the elderly (See \S \ref{subsec:concerns}). Consistent with smart home devices studies~\cite{sun2021child, pal2018analyzing}, participants prioritized child privacy and raised concerns about the usability challenges for older adults. Despite prior work suggesting that social robots function best in shared-use scenarios~\cite{Aylett_unsocial, dorrenbacher2023intricacies}, our participants were largely uncomfortable with multi-user interactions due to privacy risks. Finally, despite the AI capabilities of social robots, participants expressed relatively little concern about risks specific to generative AI. This suggest a potential lack of awareness about emerging threats, which may stem from the novelty of these devices and the tendency to associate social robots with traditional IoT devices rather than advanced, evolving, AI systems (See \S \ref{subsec:concerns}).

\boldpartitle{Social robot as an all-in-one smart home} As embodied agents, social robots could potentially replace users'  smart home devices as all-in-one home technology. Consolidating the smart home ecosystem in this way could have privacy and security implications for users. 

\emph{Increased Risk of Data Linkage.} One implication of consolidating a smart home into one device is the ease of data linkage~\cite{madaan2018data, zheng2018data}. In traditional smart homes, data collected from multiple separate IoT devices is not necessarily linked to a single user. However the multi-modal data collection of a social robot is inherently linked. This linked data can be more revealing of sensitive user information or used to infer information not explicitly provided by the user~\cite{zheng2018data}. Social robot developers should refrain from taking advantage of this linked data, as our participants expressed discomfort with data inference (See \S \ref{subsec:concerns}), and expected complete transparency on how their data would be used (See \S \ref{subsec:TransparencyExpectations}).

\emph{Mitigating Privacy Fatigue Through Centralized Management.} Privacy fatigue refers to a sense of weariness toward privacy issues, in which individual believe there is no effective way to manage their private information online~\cite{choi2018role}. This concept is closely tied to privacy resignation and cynicism, where users feel disempowered and accept pervasive data tracking as an unavoidable reality~\cite{hoffmann2016privacy}. Indeed, while almost all of our participants expressed security and privacy concerns before we explicitly raised the topic (See \S \ref{subsec:concerns}), some conveyed a sense of resignation, stating that although they were uncomfortable with data tracking, they felt it was inevitable since data collection happens across all digital devices (See \S \ref{subsec:concerns}).

As standalone devices, in lieu of an ecosystem of multiple smart devices, social robots could potentially reduce privacy fatigue by simplifying privacy management. Instead of navigating separate settings for numerous smart home devices, users could configure and monitor their privacy preferences from one place--either because the domestic social robot is the only smart device in their home or because it serves as a personal privacy assistant~\cite{ulusoy2021panola, das2018personalized} integrating with and controlling other devices. This could help users manage their privacy more efficiently, reducing their cognitive overload. If social robots are to consolidate information from across users' homes, it is imperative that they align with user security and privacy expectations and allow users fine-grained controls to specify their data-handling preferences. Our participants outlined detailed expectations regarding security and privacy measures in domestic social robots (See \S \ref{subsec: DesiredSNP}). These include the ability to review, control, and delete their information. 

\boldpartitle{Lack of regulations for social robots}
Regulations should proactively anticipate the role of domestic social robots and their impact on user privacy, including their potential function as privacy management tools. Some participants viewed policymakers as the primary stakeholders responsible for protecting consumers privacy. However, there are currently no specific, comprehensive privacy laws governing social robots. Given their extensive sensor data collection, conversational AI capabilities, and anthropomorphic design--which encourages greater user disclosure--~\cite{zhang2023s, ischen2020privacy} we argue that distinguishing social robots from standard IoT devices in legal frameworks is essential. Their ability to function as all-in-one devices further complicates regulatory considerations. For instance, one participant expressed trust in existing privacy regulations for healthcare, assuming that the Health Insurance Portability and Accountability Act (HIPAA) would apply to social robots (See \S \ref{subsec: DesiredSNP}). This is a common misconception. HIPAA only applies to ``covered entities'' such as healthcare providers and insurers, excluding non-covered entities like fitness and mental health apps. As it stands, domestic social robots do not fall under HIPAA unless they are provided by a covered entity, even if they help with health-related tasks.

Similarly, the Children's Online Privacy Protection Act (COPPA) presents limitations when applied to social robots. While COPPA restricts the collection of data \textit{from} children under 13—an issue frequently raised by our participants (See \S \ref{subsec:concerns})—it does not regulate data collected \textit{about} them. This loophole allows for the indirect collection of sensitive information \textit{about} children, undermining the law’s intended protections. In the case of social robots, a single device may be used by all household members, and, in theory, could collect data \textit{about} children present in the house through its various sensors, without necessarily collecting it directly from them.

The Federal Trade Commission (FTC) has some authority under Section 5 of the FTC Act to address unfair and deceptive trade practices. The FTC has taken action against multiple Tech companies, including a recent crackdown on deceptive AI claims~\cite{ftcAI}. For example, the FTC has taken action against DoNotPay, a company that falsely advertised an AI service as ``the first robot lawyer.''~\cite{ftcAI} However, FTC enforcement is reactive rather than proactive. It relies on consent agreements requiring companies to implement privacy and security measures for up to 20 years, with civil penalties only imposed if they violate these terms. This approach limits the FTC’s ability to deter misconduct before harm occurs. Despite some recent promising state-level efforts to regulate AI~\cite{RegAI, ColoradoAI}, privacy advocates agree that robotic technology poses challenges to existing privacy laws~\cite{calo2012robots, fosch2018did, grabler2025privacy}. AI policies that fail to consider the specific context of robotic applications risk being ineffective~\cite{fosch2018did}. As illustrated by the examples above, existing privacy regulations in the U.S. already fall short in protecting consumers from the privacy and security risks domestic social robots pose. 

Several studies have examined the extent to which robots comply with privacy regulations. Horstmann et al. identified privacy concerns in four use cases and proposed a privacy app concept to mitigate GDPR violations~\cite{horstmann2019great}. Shcafer et al. advocated for amending the UK’s Principle 4 of ethical robot design to include “transparency by design”~\cite{schafer2017spy}. Villaronga et al. explored ethical, legal, and societal challenges associated with social robots in healthcare, aligning their recommendations with recent European robotics regulatory initiatives~\cite{fosch2018did}. However, more research is needed specifically within the U.S. regulatory context. Calo laid the groundwork in 2010 by analyzing the privacy risks posed by robots in relation to U.S. regulation~\cite{calo2012robots}. However, both the robotics field and regulatory frameworks have evolved significantly since then. We urge future research to reexamine these issues, as our participants expect policymakers--alongside designers and users--to play a critical role in mitigating privacy risks.

\boldpartitle{More than just data: the overlooked privacy dimensions of domestic social robots}
There is no single definition of privacy that encompasses all of its facets. A common approach in privacy scholarship is to examine privacy through multiple dimensions, each addressing different types of privacy risks and concerns~\cite{solove2005taxonomy}. Security and privacy research in HCI has traditionally focused on informational privacy, which is central to discussions on data collection, storage, and usage--concerns that were prominent among our participants. As we elaborate throughout Section~\ref{sec:disc}, many expected clear disclosures about data handling and desired mechanisms for transparency and control. However, given the defining characteristics of social robots--embodiment, autonomy, and anthropomorphism-- it is valuable to examine their privacy implications beyond informational privacy. We discuss physical, psychological, and social privacy as key dimensions that are particularly relevant to social robots.  Notably, breaches of psychological and social privacy undermine moral-based trust, whereas breaches of informational and physical privacy tend to affect performance-based trust~\cite{callander2024navigating}. These dimensions were also examined in relation to social robots by  Callander et al., who highlighted potential infringements across different privacy types~\cite{callander2024navigating}. Below, we discuss how participants’ concerns, even when implicit, mapped onto these dimensions.

\emph{Physical Privacy.} Participants identified several threats of social robots to their physical privacy and safety (See \S \ref{subsec:concerns}). Some participants were particularly concerned about the navigation and mapping capabilities of social robots.  Given their ability to move autonomously, domestic robots may enter private spaces uninvited and create a pervasive sense of surveillance. Some participants were particularly concerned about social robots accessing children's rooms (See \S \ref{subsec:concerns}). To mitigate these risks, participants emphasized the need for mechanisms that allow users to control a social robot’s movement (See \S \ref{subsec: DesiredSNP}). Users should have the ability to lock certain rooms from the robot’s access and define boundaries within their living spaces. As Callander et al. propose, cameras on social robots should default to the off mode, reducing concerns related to unintended surveillance~\cite{callander2024navigating}. Furthermore, as expressed by participants (See \S \ref{subsec: DesiredSNP}), we suggest physical indicators, such as camera shutters or status lights, to signal when a device is actively recording audio or video.

Additionally, participants were more accepting of certain data collection practices when they aligned with their expectations (See \S \ref{subsec:concerns}). If a social robot were to intrude a user's personal space, for example to measure body temperature using its sensors through touch, transparency and explainability would be essential~\cite{callander2024navigating}. We reaffirm calls for increased transparency; social robots should clarify their capabilities, intentions, and purpose in a user-friendly manner before performing any task that might infringe upon a user's physical privacy.

\emph{Psychological Privacy.} Participants also raised concerns about how social robots `learn' and retain personal data (See \S \ref{subsec:concerns}). With the advancement and proliferation of generative AI, threats to psychological privacy are becoming increasingly relevant. Psychological privacy refers to maintaining control over information relating to our own mental states~\cite{wajnerman2021mental}. Currently, it is already possible to infer a user's mood using only data from their mobile phone sensors~\cite{Ghandeharioun}. Additionally, users prefer when conversational agents match their emotional tone~\cite{bilquise2022emotionally}. This, however, could conflict with users' privacy expectations. One participant, for example, described Google's data inference capabilities as ``creepy'' and wanted social robots to be transparent around their data inference algorithms (See \S \ref{subsec:concerns}). If social robot manufacturers want to include emotional AI features in their devices, it is imperative that they treat the inferred emotion as they would any other data. To align with participant expectations (See \S \ref{subsec: DesiredSNP}), manufacturers would need to: 1) Clearly indicate when emotional data was being inferred, 2) Allow users to review and delete the emotional data collected on them, and 3) Give users the ability to disable the emotional inference features. Beyond emotion recognition, social robots can lead to psychological dependence, chilling effects, and reduced self-reflection, particularly for vulnerable groups like children~\cite{lutz2019privacy}.

\emph{Social Privacy.} Social privacy corresponds to the social bonding and boundary management processes between humans and robots~\cite{lutz2019privacy}. Human–robot interaction studies show that people tend to anthropomorphize robots, attributing human-like qualities to them~\cite{breazeal2003toward}. This tendency can lead to increased disclosure of personal information, as users develop affection and trust toward the robot~\cite{sharkey2016should, calo2012robots, zhang2023s}. Indeed, our participants identified companionship as one of the main benefits of owning a social robot (See \S \ref{subsec:concerns}). They expected these robots to provide companionship not only for themselves but also for children and elderly individuals. Social robots' emotional expressiveness, embodiment, and communication through natural language make them uniquely well-suited to fulfill users' companionship needs. Compared to traditional IoT devices and web-based chatbots--both of which some participants already used for companionship--social robots offer a more dynamic and engaging experience, which may also feel more fulfilling.

At the same time, a notable concern emerged among participants (See \S \ref{subsec:concerns}): the potential replacement of genuine human interactions with robot-mediated ones. This concern was particularly pronounced when considering social robots for children. Some participants feared that children might prefer interacting with robots over human peers, leading to over-reliance on robotic companionship and potentially hindering their social development. In psychological therapy settings, participants worried about the loss of human touch, and feared that patients might adhere too strongly to the unreliable advice given by a social robot's language model.

To address participants' concerns on the negative impact of interacting with social robots (See \S \ref{subsec:concerns}), designers should implement safeguards that prevent excessive anthropomorphism or at least allow users to control it. Children, in particular, were seen as more vulnerable to the negative effects of interacting with social robots. Therefore, regulatory measures should be established to enforce specific design considerations for social robots marketed toward children. One option would be to design social robots to facilitate interactions \textit{between} children, rather than as a replacement of social interactions, as suggested by Aylett et al.~\cite{Aylett_unsocial}.
\section{Conclusion}
We qualitatively explored the security and privacy needs and concerns of U.S. based participants regarding domestic social robots. Our participants expressed substantial privacy and security concerns, describing a wide range of necessary measures to justify adoption. We offer actionable recommendations for developing social robots that meet the security and privacy needs of users. These include implementing clear and accessible security and privacy controls, enhancing transparency in data handling practices, and ensuring the reliability and accuracy of the robots' functionalities. 

\bibliographystyle{plain}
\bibliography{usenix2025_SOUPS}

\appendix

\section{Interview Material}\label{inteview_material}

\subsection{Interview Consent Form}\label{interview_consent}
The interview informed consent form is available at the following link:\url{https://github.com/socialrobotattitudes/SRMaterial/blob/main/Interview%20Consent%20Form.pdf}

\subsection{Screening Survey}\label{interview_screening}
The screening survey is available at the following link:\url{https://github.com/socialrobotattitudes/SRMaterial/blob/main/Screening%20Survey.pdf}

\subsection{Interview Questions}\label{interview_script}
\subsubsection{Knowledge and Awareness Toward Social Robots}

\begin{enumerate}
    \item Have you ever heard of the term social robot?
    \begin{enumerate}
        \item (\textit{If yes}) In your own words, how would you define a social robot?
        \item (\textit{If no}) If you had to guess, how would you define a social robot?
    \end{enumerate}
    \item What capabilities do you think a device should have to be counted as a social robot?
\end{enumerate}

\paragraph{Providing a definition for a social robot} While there is no single definition of what a social robot is, for the purposes of this interview, we’ll define a social robot as an artificial intelligence (AI) system that is designed to interact with humans and other robots by following social behaviors and rules attached to its role. Like other robots, a social robot is physically embodied.

\begin{enumerate}
    \item Is this definition clear to you, or is there any part of this definition that you would like us to further elaborate on?
    \item Do you know of any technologies currently available for purchase that fit the provided definition of a social robot?
\end{enumerate}

\paragraph{Robot Specification} An established company is aiming to develop a novel social robot. They are currently working on designing the prototype device, and they would like to do some research with users to specify some important aspects of the social robot before they launch the main product to the market. As an independent research institute, we are helping this company to capture people’s honest opinions and feedback regarding the prototype device to inform the future design of the company's social robot. Therefore, it is valuable for us to capture any positive and negative feedback that comes to your mind.

\paragraph{We provide the specification of the social robot} For the remainder of this interview, we may ask you a few questions to capture your attitudes and preferences toward purchasing this specific social robot, or social robots more generally. For these questions, please assume that the final price of the device is within your budget.

\begin{enumerate}
    \item Which of these features are you most comfortable with and why?
    \item Which of these features are you most concerned about and why?
    \item Would you consider purchasing this specific social robot in the near future, and why?
    \item What information would you like to have to make an informed decision as to whether or not to have this specific social robot in your home?
\end{enumerate}

\subsubsection{Contextual Privacy And Security Attitudes}
In this section, we are going to walk through some potential use-case scenarios for the prototype social robot we described earlier. We will ask some follow-up questions after each scenario.

[\textbf{Purchasing for yourself}] Imagine that you are living alone, and you are purchasing this specific social robot for yourself to be the robot’s primary user. 

[\textbf{Purchasing for children}] Imagine that you are living in a family setting with a child, and you are purchasing this specific social robot for the child to be the robot’s primary user. 
 
[\textbf{Purchasing for the elderly}] Imagine that you are living with an elderly family member, and you are purchasing this specific social robot for the elderly family member to be the robot’s primary user. 

[\textbf{Purchasing for a communal household}] Imagine that you are living in a communal setting, and you are purchasing this specific social robot to be shared amongst your household without having a specific primary user.

So far we have talked about purchasing the device for yourself, a child, or an elderly family member. Now we’re going to talk about the various contexts that this device could be used in.

[\textbf{Educational Context}] Keeping in mind the four user scenarios we just described, how comfortable or concerned would you be purchasing this device to fulfill an educational need for the users?

[\textbf{Medical Context}] Keeping in mind the four user scenarios we just described, how comfortable or concerned would you be purchasing this device to fulfill a medical need for the users?

[\textbf{Psychological Therapy Context}] Keeping in mind the four user scenarios we just described, how comfortable or concerned would you be purchasing this device to fulfill a psychological therapy need for the users?

\begin{enumerate}
    \item How comfortable or concerned are you with this described scenario, and why?
    \begin{enumerate}
        \item (\textit{If concerned}) What do you think should happen to make you less concerned about this described scenario?
    \end{enumerate}
\end{enumerate}



\subsubsection{Privacy And Security Expectations Toward Social Robots}

\begin{enumerate}
    \item On a scale of 1 to 5, 1 being not at all important and 5 being very important, how important do you consider privacy and security to be in your decision to purchase a social robot and why?
    \item What type of privacy and security information, if any, would you want to know about to determine if you would purchase a social robot?
    \item As a reminder, social robots collect information, and use artificial intelligence to enable social interactions. What type of information, if any, would you want to know about the conversational artificial intelligence features of a social robot to determine if you would purchase it?
    \item How would you like this information to be communicated to you to inform your purchasing decisions?
    \item Who do you think is responsible for protecting users from the potential privacy and security risks of social robots, and how?
\end{enumerate}

\begin{table*}[htbp]
\centering
\resizebox{\textwidth}{!}{%
\footnotesize{
\begin{tabular}{ll}
\toprule[1.1pt]
\textbf{Feature} & \textbf{Description} \\ \midrule
Visual Recognition & \makecell[l]{The robot can detect and remember different faces and objects in its environment. In addition, it can react and respond \\to the visual cues.} \\
Voice Recognition & \makecell[l]{The robot can detect and remember the different voices in its environment. In addition, it can react and respond to \\voice commands.}\\
Expressive Communication & The robot can communicate via expressions. Such expressions could be communicated through voice or facial expressions.\\
Personalization & The robot learns to personalize its interactions with its users. It will adapt to your preferences and behaviors. \\
Navigation and Mapping & The robot can map its environment and navigate through spaces without collision.\\
Internet Connected & The device is connected to the internet, allowing for app downloads.\\
\bottomrule[1.1pt]
\end{tabular}}}
\vspace{-.15in}
\caption{We designed a specification for the social robot based on the prominent capabilities in existing social robots.}
\label{table:SRSpecification}\vspace{-.15in}
\end{table*}

\begin{table*}[t]
\centering
\resizebox{\textwidth}{!}{%
\footnotesize{
\begin{tabular}{cp{12cm}p{8cm}}
\toprule[1.1pt]
\textbf{ID} & \textbf{Smart Home Devices} & \textbf{AI-Enabled Chatbots} \\ \midrule
P1 & Activity tracker, Home assistants, Connected printers, Smart doorlock & ChatGPT, Google Bard, Google Gemini \\
P2 & Smart TV, Smart thermostat, Connected lights, Games console, Home assistants, Video streaming product, Smart plugs, Smart doorlock & ChatGPT, Google Bard, Google Gemini, Microsoft Bing AI \\
P3 & Smart TV, Activity tracker, Smart thermostat, Connected lights, Games console, Home assistants, Smartwatch, Video streaming product, Connected printers, Smart plugs, Smart doorlock, Baby camera, Smart kitchen appliances & ChatGPT, Google Bard, Google Gemini, Snapchat My AI \\
P4 & Smart TV, Connected lights, Games console, Home assistants, Smartwatch, Video streaming product, Connected printers, Smart plugs, Smart doorlock & ChatGPT, Google Bard, Google Gemini, Microsoft Bing AI \\
P5 & Home assistants & ChatGPT \\
P6 & Smart TV, Activity tracker, Smart thermostat, Connected lights, Games console, Home assistants, Smartwatch, Video streaming product, Connected printers, Smart plugs, Smart doorlock, Smart water sprinkler, Smart kitchen appliances & ChatGPT \\
P7 & Smart TV, Games console, Home assistants, Smartwatch, Video streaming product, Connected printers & ChatGPT \\
P8 & Smart TV, Connected lights, Home assistants, Video streaming products, Connected printers, Smart plugs, Smart kitchen appliances & ChatGPT, GitHub Copilot, Google Bard, Google Gemini, Microsoft Bing AI \\
P9 & Smart TV, Activity tracker, Connected lights, Games console, Home assistants, Video streaming product, Connected printers, Smart plugs, Smart security camera & ChatGPT, Google Bard, Microsoft Bing AI \\
P10 & Smart TV, Activity tracker, Connected lights, Home assistants, Smart watch, Video streaming product, Connected printers, Smart plugs, Smart security camera, Smart health monitors, Smart kitchen appliances, Smart bluetooth trackers & ChatGPT, Google Bard \\
P11 & Smart TV, Games console, Home assistants, Video streaming product, Connected printers & ChatGPT, Google Bard, Snapchat My AI \\
P12 & Smart TV, Activity tracker, Smart thermostat, Games console, Home assistants, Connected printers, Smart health monitors & ChatGPT, Google Bard, Microsoft Bing AI \\
P13 & Smart TV, Activity tracker, Games console, Home assistants, Video streaming product, Connected printers, Smart security camera, Grocery ordering, Smart kitchen appliances & ChatGPT, Claude, Google Bard, Google Gemini, Jasper, Microsoft Bing AI \\
P14 & Smart TV, Games console, Home assistants, Smartwatch, Video streaming product, Connected printers, Smart water sprinkler & ChatGPT, Claude, Google Bard, Google Gemini \\
P15 & Smart TV, Activity tracker, Smart thermostat, Connected lights, Home assistants, Smartwatch, Connected printers, Smart plugs, Smart doorlock, Smart security camera, Smart smoke monitors & ChatGPT, Google Bard \\
P16 & Home assistants, Video streaming products, Connected printers, Smart doorlock, Smart security camera, Baby camera, Smart kitchen appliances & ChatGPT, Google Bard, Microsoft Bing AI \\
P17 & Smart TV, Activity tracker, Connected lights, Games console, Home assistants, Video streaming product, Connected printers, Smart plugs, Smart security camera, Smart health monitors, Smart kitchen appliances & ChatGPT, Google Bard, Microsoft Bing AI, Snapchat My AI, DeepAI \\
P18 & Smart TV, Connected lights, Games console, Home assistants, Smartwatch, Video streaming product, Connected printers, Smart security camera & ChatGPT, Claude \\
P19 & Activity tracker, Games console, Home assistants, Video streaming product, Connected printers, Smart doorlock, Smart security camera & MetaAI \\
\bottomrule[1.1pt]
\end{tabular}
}
}
\caption{Participants and the smart home devices and AI-enabled chatbots they use.}
\label{tab:technologies}
\end{table*}

\subsection{Debriefing Statement}\label{interview_debrief}
The debriefing statement is available at the following link:

\url{https://github.com/socialrobotattitudes/SRMaterial/blob/main/Debriefing%20Statement.pdf}

\section{Interview Codebook}\label{interview_codebook}
The codebook is available at the following link:

\url{https://github.com/socialrobotattitudes/SRMaterial/blob/main/Codebook%20(F).pdf}

\end{document}